 \pdfoutput=1
\documentclass[12pt]{article} 

\usepackage[T2A]{fontenc}
\usepackage[utf8]{inputenc}
\usepackage{geometry}
\usepackage[english]{babel}
\usepackage{graphicx} 

\usepackage{float} 

\graphicspath{{figures/}} 

\begin{document}



\newcommand{\HRule}{\rule{\linewidth}{0.5mm}} 

{\center 


{ \huge \bfseries Alignment and resolution studies\\ of a MARS scanner}\\[0.4cm] 


A.~Gongadze, A.~Zhemchugov, G.~Chelkov,  D.~Kozhevnikov, I.~Potrap, M.~Demichev, P.~Smolyanskiy,  R.~Abramishvili, S.~Kotov, A.P.~Butler$^1$, P.H.~Butler$^2$, S.T.~Bell$^3$\\

{\small \it {$^a$}University of Otago, Christchurch, NZ}\\
{\small \it {$^b$}University of Canterbury, Christchurch, NZ}\\
{\small \it {$^c$}MARS Bioengineering, 29a Clyde Road, Christchurch, NZ}\\
{\small \it {$^d$}Joint Institute For Nuclear Research, Dubna}










}
\section{Introduction}

The MARS scanner\cite{mars} is designed for the x-ray spectroscopic study of samples with the aid of computer tomography methods.
Computer tomography allows the reconstruction of slices of an investigated sample using a set of shadow projections obtained for different angles.
Projections in the MARS scanner are produced using a cone x-ray beam geometry. Correct reconstruction in this scheme requires precise knowledge of several geometrical parameters of a tomograph, such as displacement of a rotation axis, x-ray source position with respect to a camera, and camera inclinations. Use of inaccurate parameters leads to a poor sample reconstruction. Non-ideal positioning of camera, x-ray source and cylindrical rotating frame (gantry) itself on which these parts are located, leads to the need for tomograph alignment. In this note we describe the alignment procedure that was used to get different geometrical corrections for the reconstruction. Also, several different estimations of the final spatial resolution for reconstructed images are presented.

\section{The MARS scanner}

\begin{figure}[H]
\center{\includegraphics[width=0.95\linewidth]{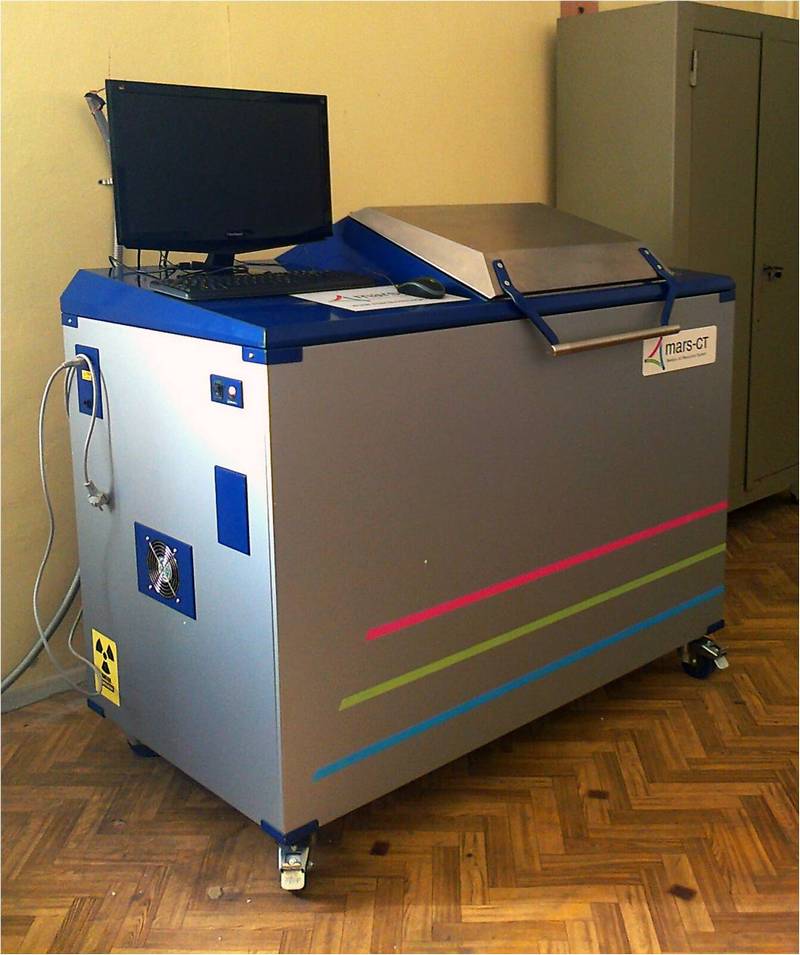}}
\caption{The MARS scanner in the Laboratory of Nuclear Problems of the Joint Institute for Nuclear Research.}
\label{fig:mars}
\end{figure}

The MARS (Medipix All Resolution System) scanner (Figure~\ref{fig:mars}) obtained by the Laboratory of Nuclear Problems of the Joint Institute for Nuclear Research, has a camera with two detectors based on gallium arsenide (compensated with chromium) sensors\cite{GaAs} and equipped with the Medipix3.1\cite{medipix} electronics. Each detector has 256 x 256 pixels of 55 microns and has dimensions of 14 x 14 mm.
The Medipix3.1 electronics can simultaneously count the number of photons with energies above the threshold at each pixel for two different thresholds. Gallium arsenide sensors have higher detection efficiency of photons at high energies in comparison with silicon sensors. The advantages of the scheme with single photon counting include the absence of dark currents and radiation dose reduction. Moreover, by changing user-settable parameters in the Medipix3.1 electronics, it is possible to select different ranges of photon energies for spectrometry.

\begin{figure}[H]
\center{\includegraphics[width=0.95\linewidth]{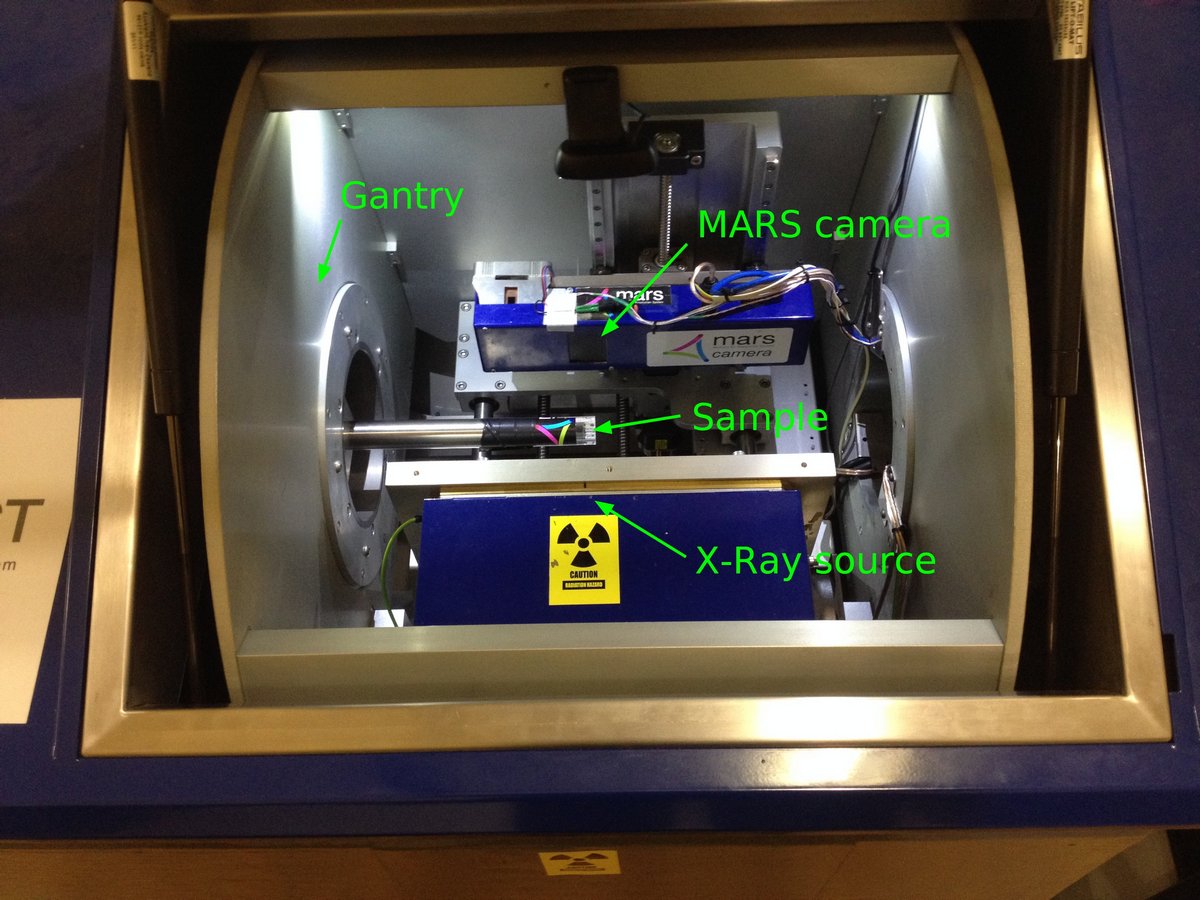}}
\caption{Photo of the MARS equipment with a scanned sample.}
\label{fig:mars_gentry}
\end{figure}

Since the MARS tomograph was designed for biomedical research, the gantry with the scanning equipment (x-ray source and camera) attached to it is rotated around fixed scanned sample~(Figure~\ref{fig:mars_gentry}). The gantry rotation axis is horizontal. A test sample (up to 100~mm in diameter and 300~mm length) is placed in the center and can be moved along the rotation axis. The camera and the source are placed on the gantry from the opposite sides and are able to move along the axis connecting each other, approaching or moving away from the sample. In order to enable scanning of large samples camera can also be moved in the direction transverse to the rotation axis and the axis connecting the source and the camera.

\subsection{The MARS tomograph reference frame}

\begin{figure}[H] 
\center{\includegraphics[width=0.95\linewidth]{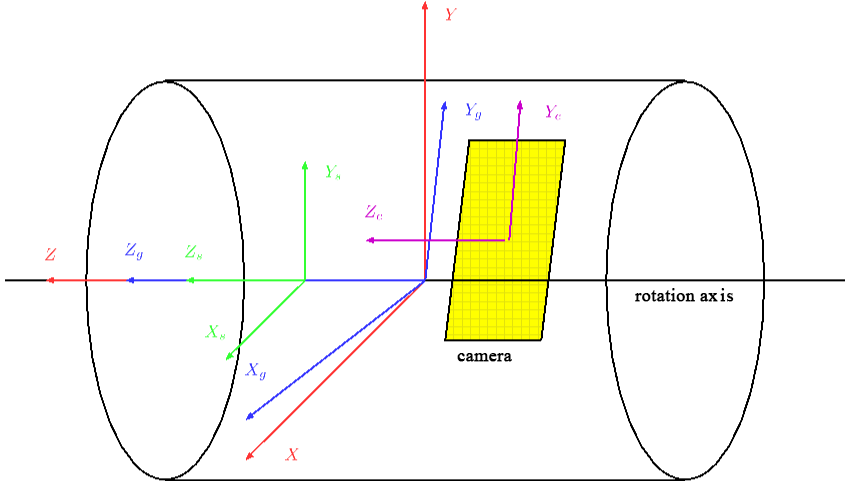}}
\caption{Reference frames associated with the moving parts of the tomograph} 
\label{fig:coorsys}
\end{figure}

To describe the geometry of the tomograph it is necessary to define the reference frames associated with the moving parts of the tomograph (Figure~\ref{fig:coorsys}):

\begin{description}

\item{Laboratory reference frame} \hfill \\
The Cartesian reference frame ($X$, $Y$, $Z$), associated with the tomograph support structure. The axes are as it is shown in the figure.

\item{Sample reference frame} \hfill \\
The Cartesian reference frame ($X_s$, $Y_s$, $Z_s$), associated with the sample, which can be moved along the $ Z $ axis of the laboratory reference frame.

\item{Gantry reference frame} \hfill \\
The Cartesian reference frame ($X_g$, $Y_g$, $Z_g$), associated with the gantry, and differs from the laboratory frame by the rotation angle $\phi$ around the $Z$ axis.

\item{Camera reference frame} \hfill \\
The Cartesian reference frame $X_c$, $Y_c$, $Z_c$), associated with the camera that can be moved along the $Y_g$ axis of the gantry reference frame. This reference frame is defined by the pixel detector plane. The $Y_c$ axis is parallel to the detector rows of pixels, the $Z_c$ axis is parallel to the detector columns of pixels.

\end{description} 

\section{Mechanical precision of the movement of the tomograph parts}

\subsection{Measurements using an electronic feeler gauge}

\begin{figure}[H]
\center{\includegraphics[width=0.95\linewidth]{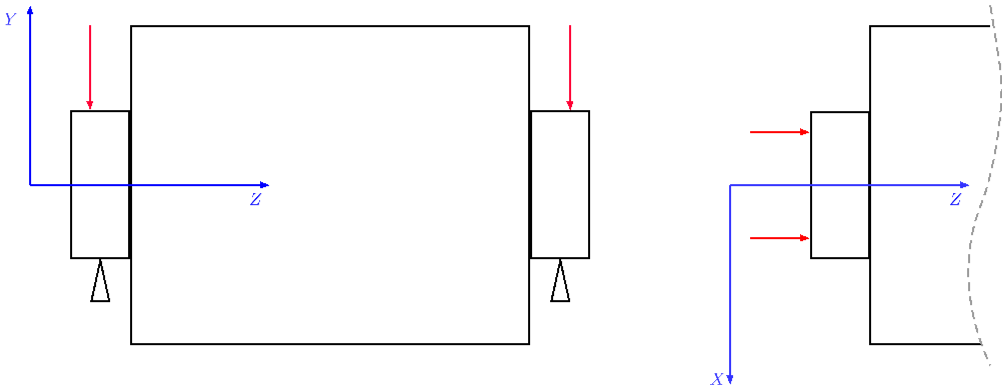}}
\caption{Measurement of the rotation axis Z displacement along vertical Y-axis (left) and measurement of the gantry movement along the Z-axis (right).}
\label{fig:gentry}
\end{figure}

Prior to the alignment studies, the mechanical precision of the movements of the tomograph parts has been investigated. The main goal was to measure the displacement of the gantry rotation axis during its rotation and to measure the displacement of the gantry itself along the rotation axis. To perform these measurements an electronic feeler gauge was used. In the first case, the positions of the upper points of the left and right gantry supports (red arrows in Figure~\ref{fig:gentry}~left side) have been measured for different rotation angles. The mean value of these two measurements (left and right) gives an estimation of the displacement of the rotation axis in the local area of a measured sample.
In the second case, position measurements have been performed from the near side and the far side of the edge of the left gantry support (red arrows in Figure~\ref{fig:gentry}~right side). In this case the mean value of the two measurements describes gantry movement along its rotation axis.

\begin{figure}[H] 
\center{\includegraphics[width=0.95\linewidth]{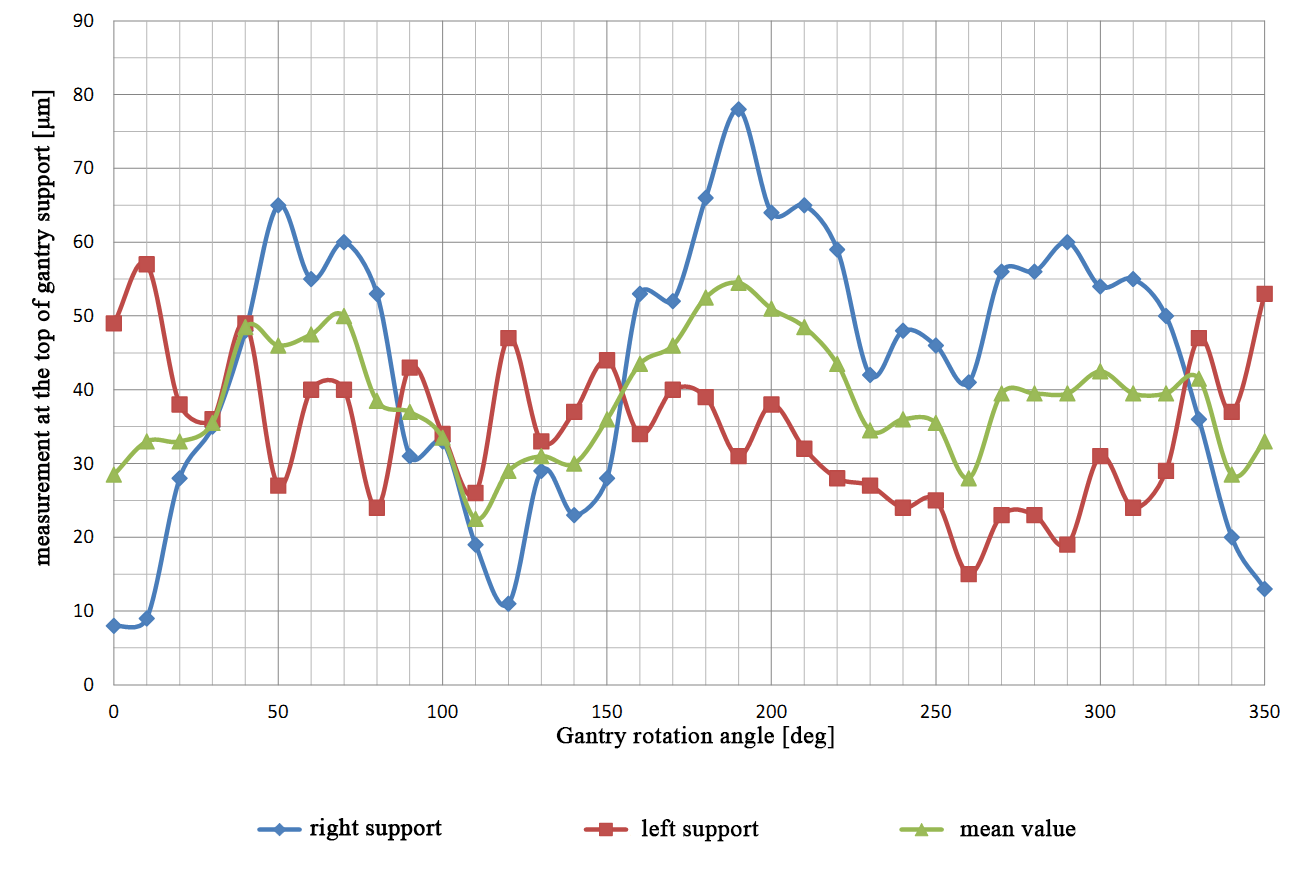}}
\caption{Relative position of upper points of the left (red) and right (blue) gantry supports as a function of the gantry rotation angle. The mean value (green) characterizes the displacement of the gentry rotation axis in the scanning area during rotation.}
\label{fig:backlash_y}
\end{figure}

\begin{figure}[H] 
\center{\includegraphics[width=0.95\linewidth]{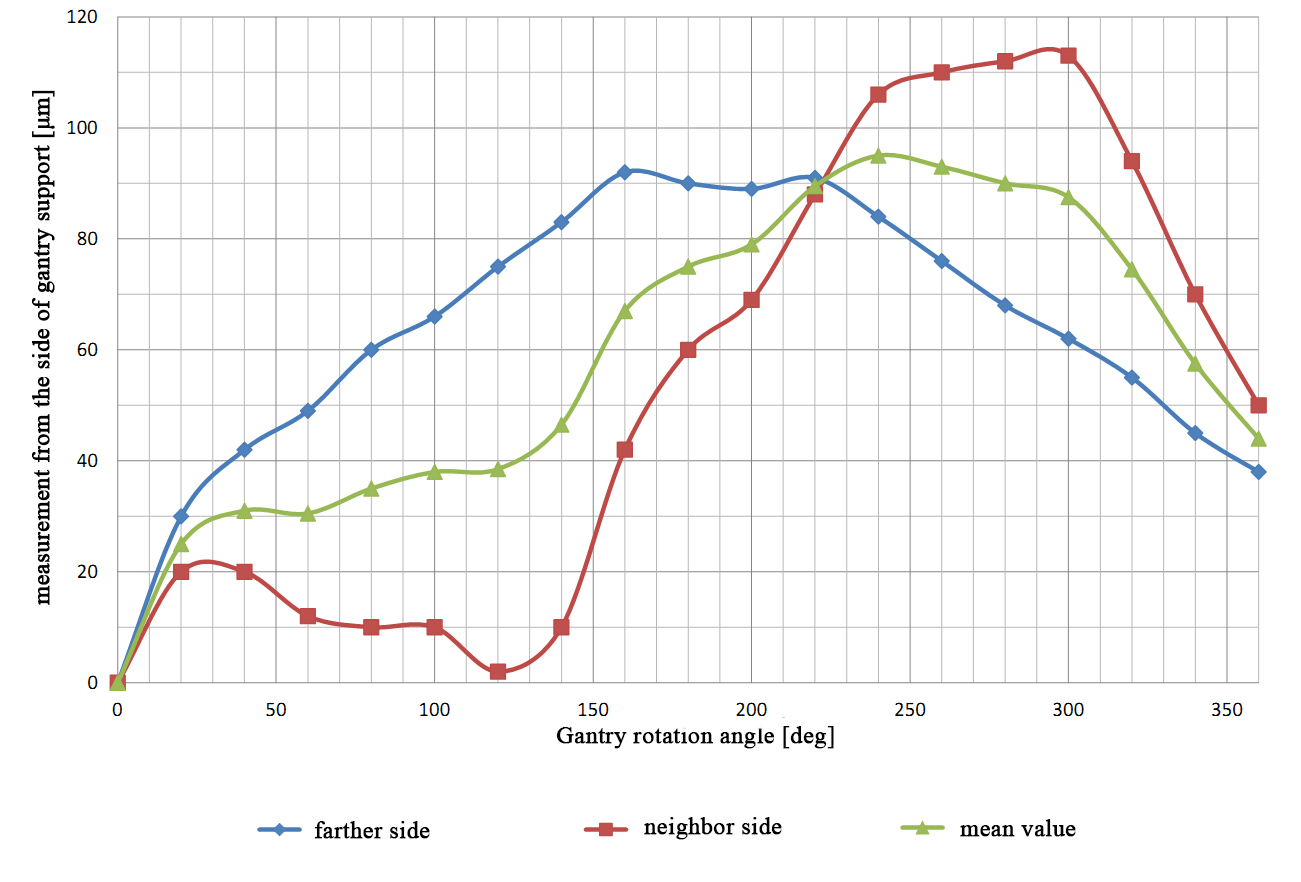}}
\caption{Position measurements of the "near" (red) and "far" (blue) sides of the left gantry support for different rotation angles. The mean value (green) characterizes the displacement of gantry along its rotation axis.}
\label{fig:backlash_z}
\end{figure}

\vspace {5mm}
The results of the described mechanical measurements are presented in Figure~\ref{fig:backlash_y} and Figure~\ref{fig:backlash_z} as a function of the gantry rotation angle. The results show that in the area of a scanned object the displacement of the gantry rotation axis is withing 30 microns (green line in Figure~\ref{fig:backlash_y}), while the gantry itself moves along Z-axis by 100 microns (green line in Figure~\ref{fig:backlash_z}). These displacement values are not negligible with respect to the pixel size of the detector and will lead to additional smearing of a reconstructed image. The results of these measurements may be used in the future for correcting projections of scan depending on the angle of rotation.

\subsection{Measurements using angular encoder system}

Rotation of the gantry was studied using the optical angular encoder system RENISHAW RESR\cite{resr}.
The system we used consists of the RESR ring (with the diameter of 200~mm) with graduations marked on it every 20~$\mu$m and the optical  readhead with a resolution of 5~$\mu$m. The RESR ring is mounted on the shaft of the gantry and the readhead is attached to the tomograph frame. The angular resolution of this system is about 10 arc seconds.

\begin{figure}[H]
\center{\includegraphics[width=0.95\linewidth]{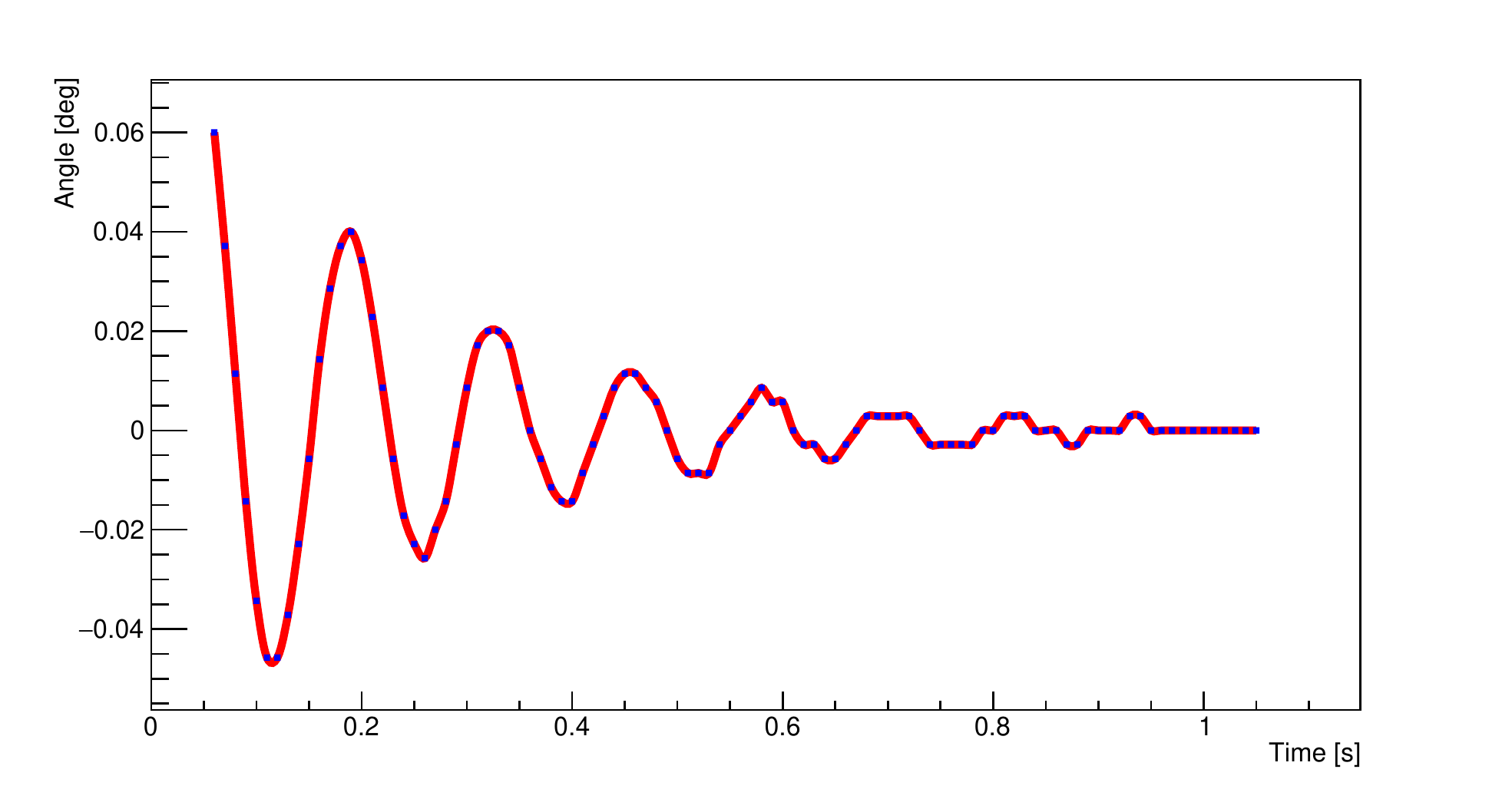}}
\caption{Angular measurements after stopping the tomograph rotation.}
\label{fig:vibrations}
\end{figure}

In the normal scanning procedure the tomograph equipment rotates around a fixed sample making stops to get the x-ray projection images at different rotation angles. The gantry with its equipment attached has a considerable moment of inertia, which leads to vibrations of tomograph parts at the moment it stops. The behavior of the tomograph at this moment was investigated with the aid of RESR system (Figure~\ref{fig:vibrations}). As shown in  the figure, significant initial vibrations of the system are observed. To get projections of good quality, there should be a time delay between the stop and the exposure of about half of a second.

\begin{figure}[H]
\center{\includegraphics[width=0.95\linewidth]{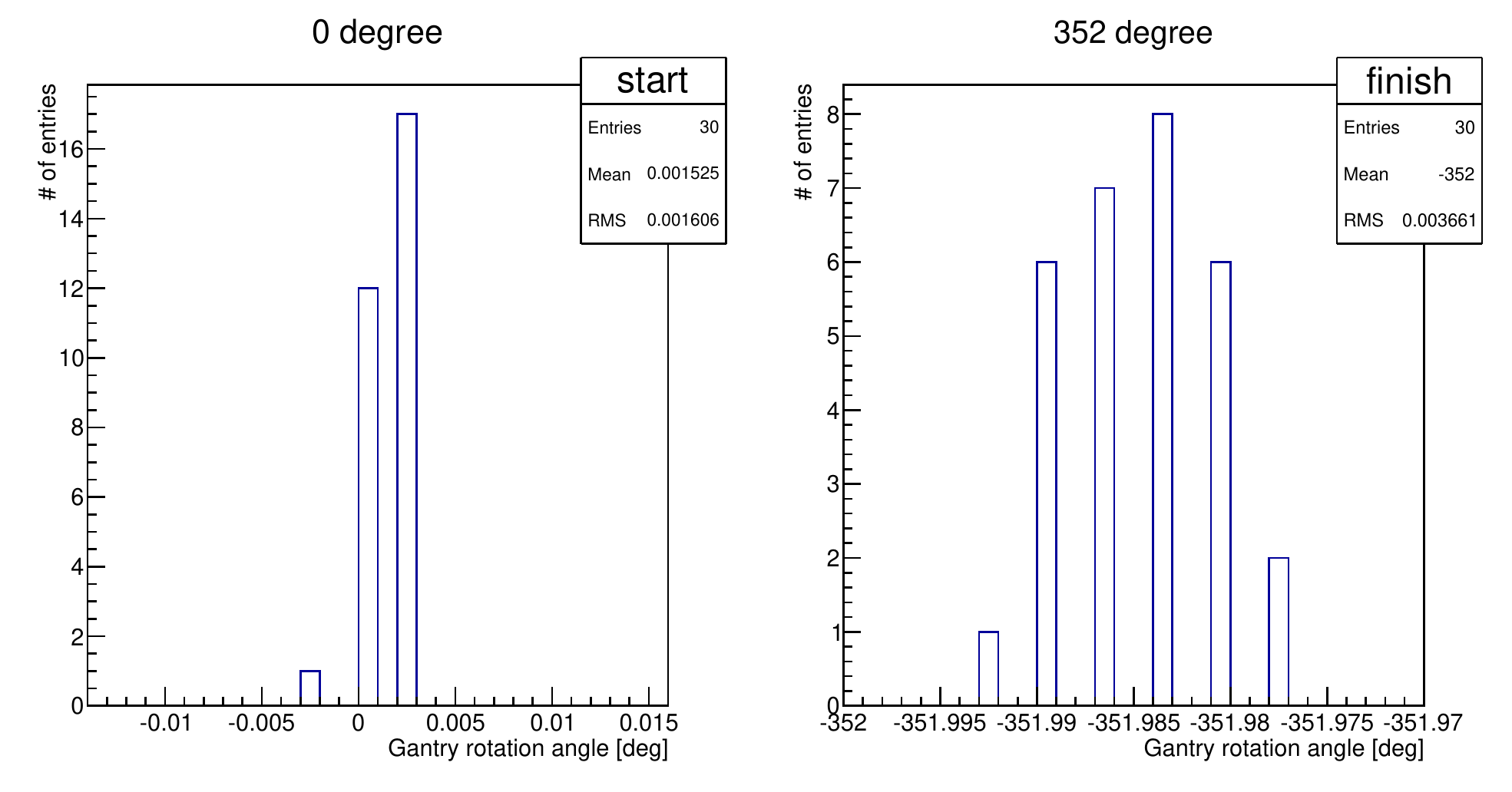}}
\caption{Repeatability of angular measurements when tomograph is moved to positions 0$^\circ$ and 352$^\circ$.}
\label{fig:repit}
\end{figure}

The next step of the investigation is to study  the repeatability of angular measurements at the same positions of the step motor which rotates the gantry.
During a scan the tomograph makes stops several times in the same angular positions to get projections. For successful stitching of these projections, a high accuracy in the repeatability of angular positions is needed. Figure~\ref{fig:repit} shows the results of numerous angular measurements when tomograph is moved to positions 0$^\circ$ and 352$^\circ$. The results show that the design of the tomograph does provide high reproducibility of angular positions.

\begin{figure}[H]
\center{\includegraphics[width=0.95\linewidth]{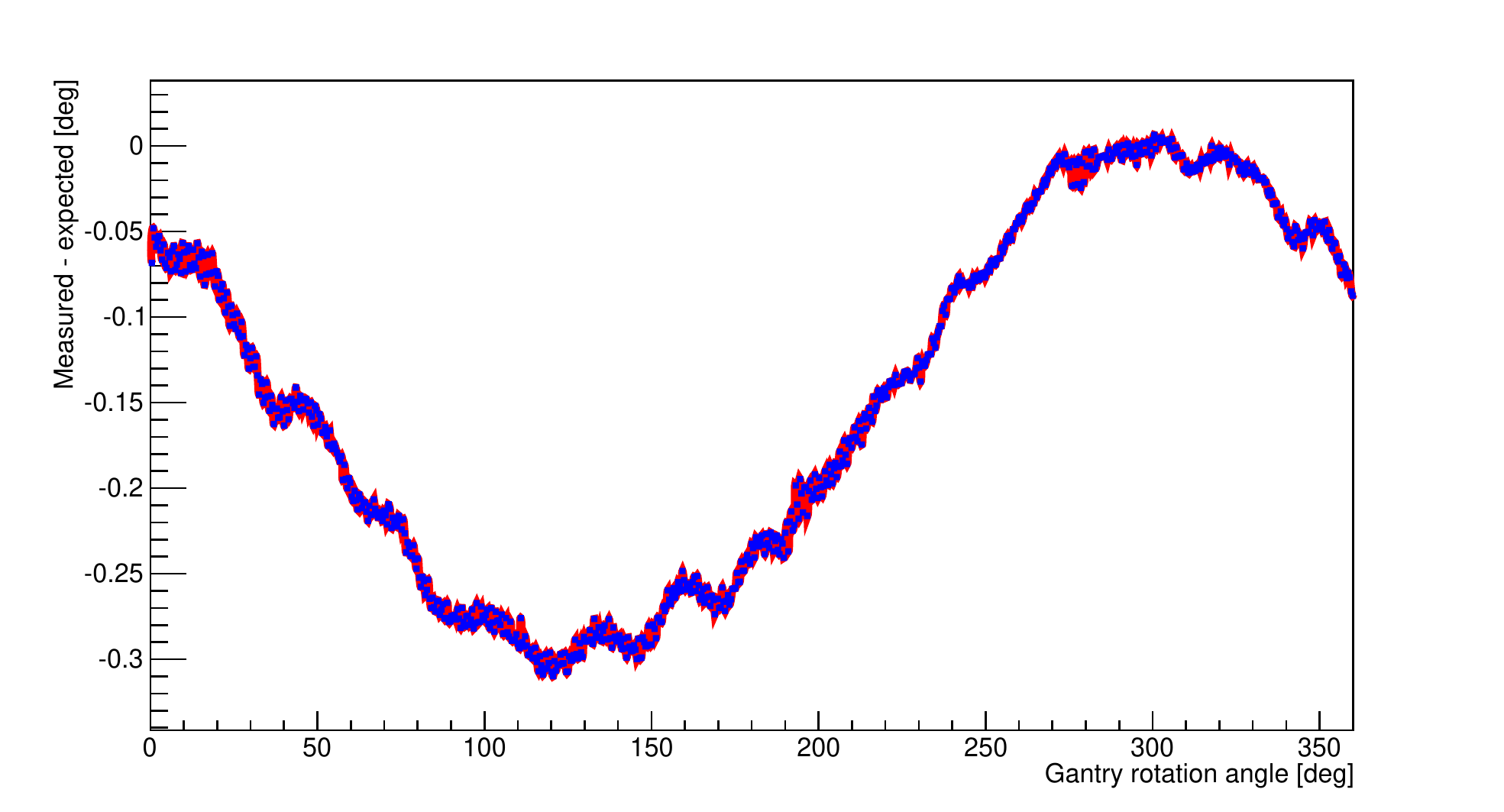}}
\caption{The difference between measured and expected values of angular positions as a function of gantry rotation angle.}
\label{fig:gantryAngle}
\end{figure}

The gantry of the MARS scanner is driven by a step motor via an automotive timing belt. The weight distribution of the equipment attached to the gantry is non-uniform and can lead to different tension of the belt for different rotation positions. The following measurements have been performed to check whether the tomograph angular positions correspond to actual angular positions. In Figure~\ref{fig:gantryAngle} the difference between measured and expected values of angular positions is presented as a function of gantry rotation angle.
Deviation of the measured angle is significant and its maximum approximately corresponds to the value of 500 microns on the surface of the RESR ring. In principle, encoder measurements are affected not only by the ring rotation but also by the movement of the ring axis during its rotation. To account for this effect,
independent measurements of the ring axis movement were performed depending on the rotation angle. The distance to the surface of the RESR ring was measured using SAFIBRA optical probe. The amplitude of this measurement does not exceed 80 microns and can not explain the value of the difference in measured rotation angle.
The difference between the measured and expected values of the rotation angle can be taken into account to correct gantry positions and then used in the image reconstruction procedure with nonlinear rotation steps.

\section{Geometrical parameters requiring calibration}

In case of ideal geometry, the $Z_s$ axis of the MARS tomograph, along which the sample is moving, is parallel to the gantry rotation axis $Z_g$, parallel to the columns of the camera $Z_c$ axis and parallel to the $Z$ axis of the laboratory reference frame.
The $Y_c$ and $Z_c$ axes corresponding to the camera pixel rows and pixel columns are parallel to the $Y_g$ and $Z_g$ axes, respectively. 
The axis along which the x-ray source and the camera are moving in the direction toward the sample or from the sample, coincides with the $X_g$ axis.
The focus of the x-ray source and the center of the camera are also lying on this axis, and it is crossing with the gantry rotation axis. The distances from this crossing point to the source SOD, the crossing point to the camera ODD, and from the source to the camera SDD are assumed to be well determined. The $Y_c$ axis $Y_c$, along which the camera moves is parallel to the $Y_g$ axis. 

In reality, the geometry has some misalignment which can be expressed in small corrections on shifts and rotations of the tomograph parts.
Some of these geometrical corrections can be taken into account in the reconstruction of three-dimensional image of a sample, so it is necessary to know their values accurately.
One of these parameters is the camera inclination angle with respect to the rotation axis, i.e. the $\alpha_{\tt{tilt}}$ angle between the $Y_c$ axis and the $Y_g$ axis. We also need to take into account the position of the projection of the gantry rotation axis on the plane of the camera. It is also important to know the position of the projection of the source on the camera plane along the $Y_c$ and $Z_c$ axes. Finally, one should accurately measure the distance from the source to the rotation axis SOD and from the rotation axis to the camera ODD.

This work is devoted to finding the camera inclination angle $\alpha_{\tt{tilt}}$, the determination of the position of the source projection on the camera plane, and the refinement of the distance between the source and the camera SDD. The position of the projection of the rotation axis on the camera plane is computed by iterative methods with good precision.

\section{The procedure for the MARS scanner alignment}

\subsection{Phantom description}

\begin{figure}[H] 
\center{\includegraphics[width=0.8\linewidth]{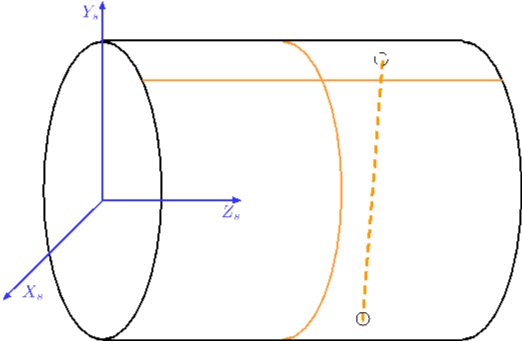}}
\caption{Phantom for the tomograph geometry calibration}
\label{fig:fantom}
\end{figure}

To study the geometry of the tomograph a very simple phantom was designed. This is a plastic hollow cylinder of 35~mm diameter with two tungsten 50~$\mu$m wires glued on the surface of the cylinder (Figure~\ref{fig:fantom}). The tungsten wires form a "cross" with one wire oriented along the Z-axis and the other in the perpendicular direction. 

\begin{figure}[H] 
\center{\includegraphics[width=0.95\linewidth]{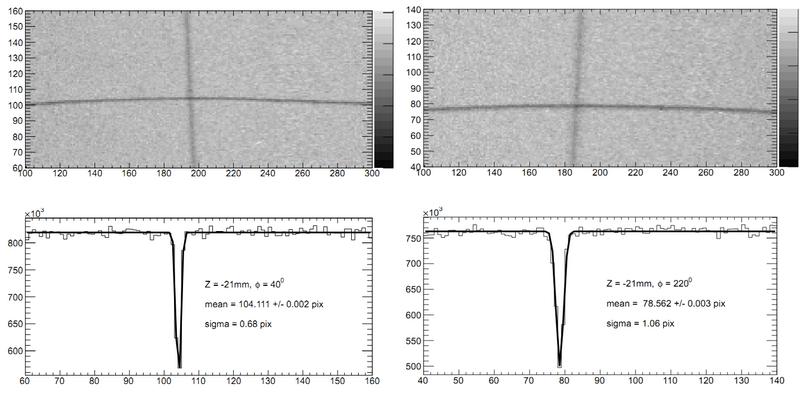}}
\caption{Determination of the position of the shadow cross, using a Gaussian approximation in the local vicinity of a crossing point.}
\label{fig:fantom2}
\end{figure}

On the camera projections, the position of the cross can be determined using a 
Gaussian approximation of the shadow in the local vicinity of a crossing point. An example of these measurements is presented in Figure~\ref{fig:fantom2}. The position of the shadow cross was determined for the holder position -21~mm, and the gantry rotation angle 40$^\circ$ in the first case (left side) and also for the gantry angle 220$^\circ$ in the second case (right side).

The above mentioned cross of wires, glued to the surface of the cylinder, was mainly used in the procedure for determining the source position. To find the camera inclination angle, another wire was stretched in the cross section of the phantom, which is shown as a vertical dashed line in Figure~\ref{fig:fantom}.

\subsection{Determination of the camera inclination angle $\alpha_{\tt{tilt}}$.}

\begin{figure}[H] 
\center{\includegraphics[width=0.65\linewidth]{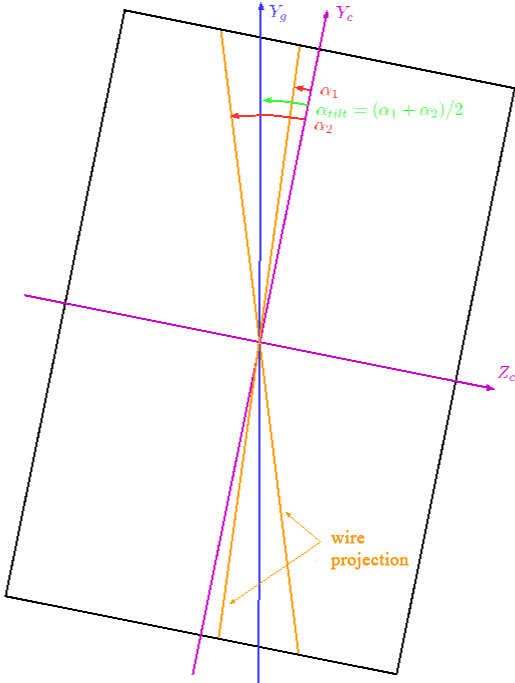}}
\caption{Determination of the camera inclination angle $\alpha_{\tt{tilt}}$ with respect to the gantry rotation axis in the plane $Y_gZ_g$ plane.}
\label{fig:calib3}
\end{figure}

To determine the camera inclination angle $\alpha_{\tt{tilt}}$ in the $Y_gZ_g$ plane, the projection of the wire stretched in the $X_sY_s$ plane is used. For this purpose the shadow inclination angle is analyzed for the vertical wire. Because the wire itself is not perfectly vertical, two projections are used: direct projection, and one obtained by rotating the gantry 180 $^\circ$. The average angle of the two positions gives the desired camera inclination angle in the plane $Y_gZ_g$ (Figure~\ref{fig:calib3}).

\subsection{Determination of the source position along the $Z_c$ axis}

\begin{figure}[H] 
\center{\includegraphics[width=0.95\linewidth]{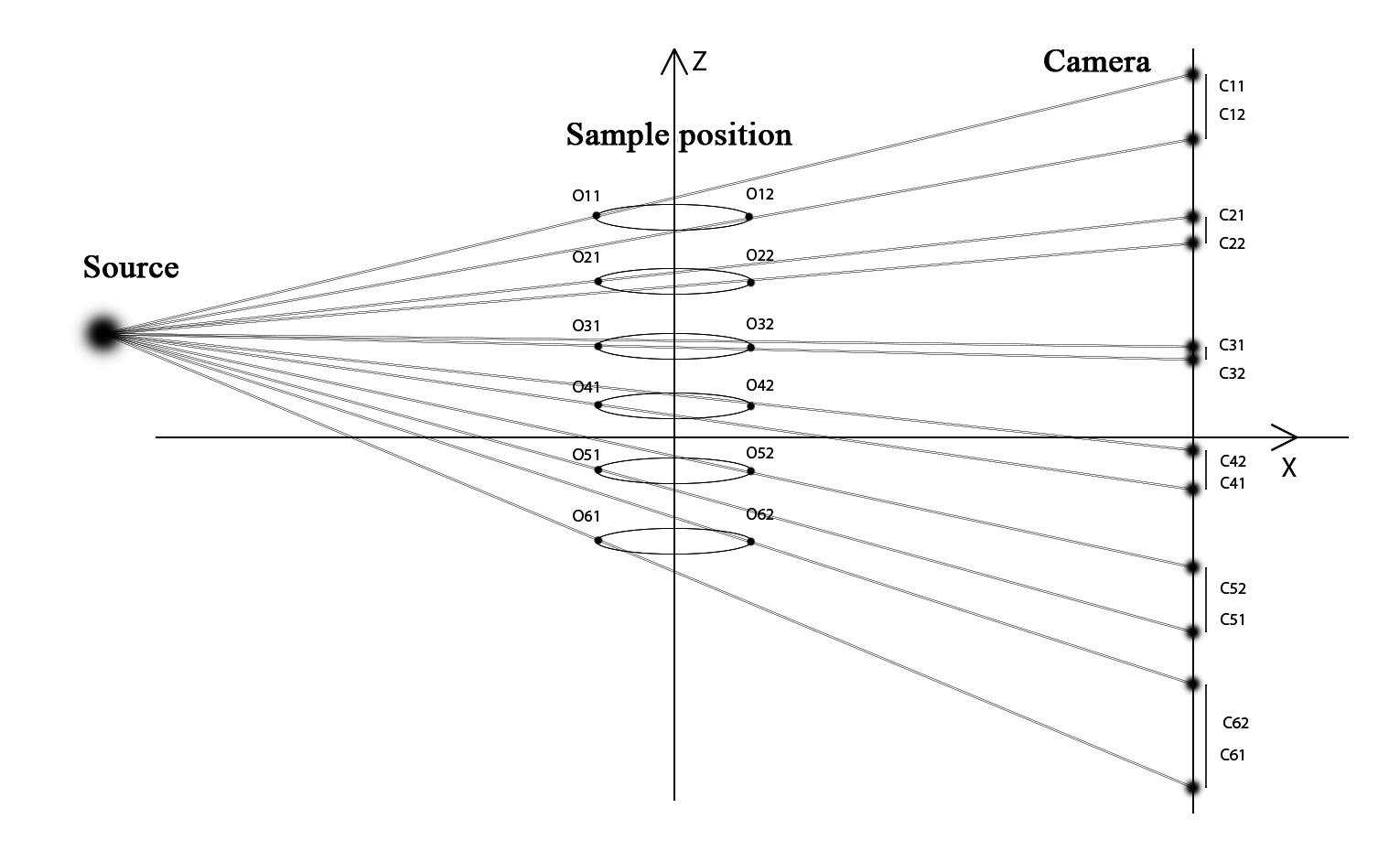}}
\caption{The procedure for the source $Z$-position determination.}
\label{fig:calib2}
\end{figure}

To determine the source position with respect to the camera along the $Z_c$ axis, we used the shadow cross images for the two gantry rotation angles in six $Z$-positions of the sample holder (Figure~\ref{fig:calib2}). The values of rotation angles of 40$^\circ$ and 220$^\circ$ have been chosen that correspond to the "near" and "far" positions of the cross with respect to the camera. In principle, it is not necessary for the angle difference to be 180$^\circ$ and several rotation positions can be used instead of two. 

\begin{figure}[H]
\center{\includegraphics[width=0.95\linewidth]{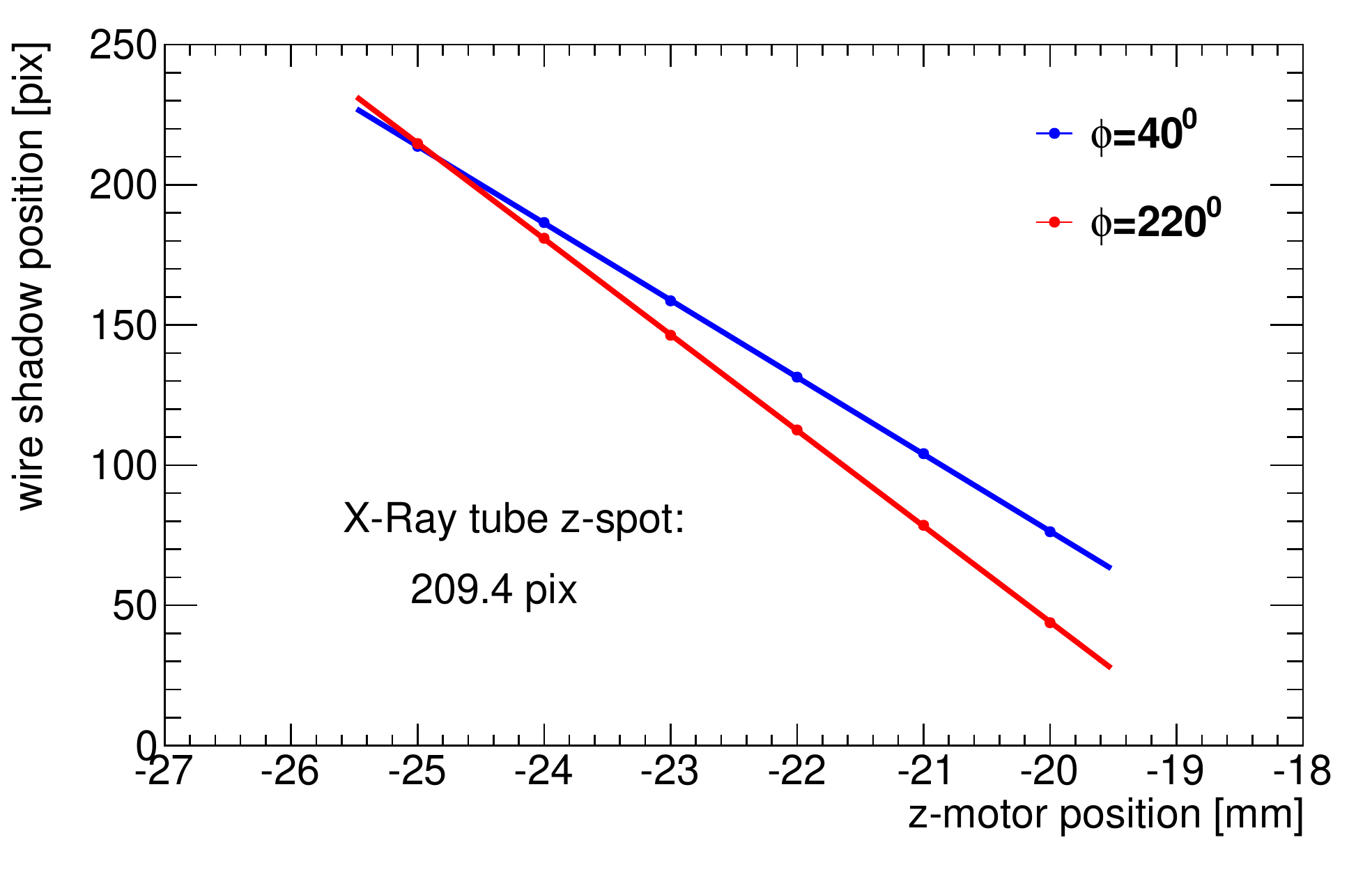}}
\caption{Determination of the source position along the $Z_c$ axis}
\label{fig:calib_res2}
\end{figure}

For each of the projections, the position of the shadow crossing point along the camera $Z_c$ axis has been calculated, as it is shown in Figure~\ref{fig:fantom2}. The dependence of the shadow cross position on the phantom $Z$-position is presented in Figure~\ref{fig:calib_res2}. The blue line corresponds to the measurements  at 40$^\circ$ rotation angle and the red line corresponds to the measurements  at 220$^\circ$ rotation angle. It is obvious that the crossing point of these two lines is the desired source position along the camera $Z_c$ axis in pixels.

\subsection{Determination of the source position along the $Y_c$ axis and of the source to detector distance SDD}

\begin{figure}[H] 
\center{\includegraphics[width=0.95\linewidth]{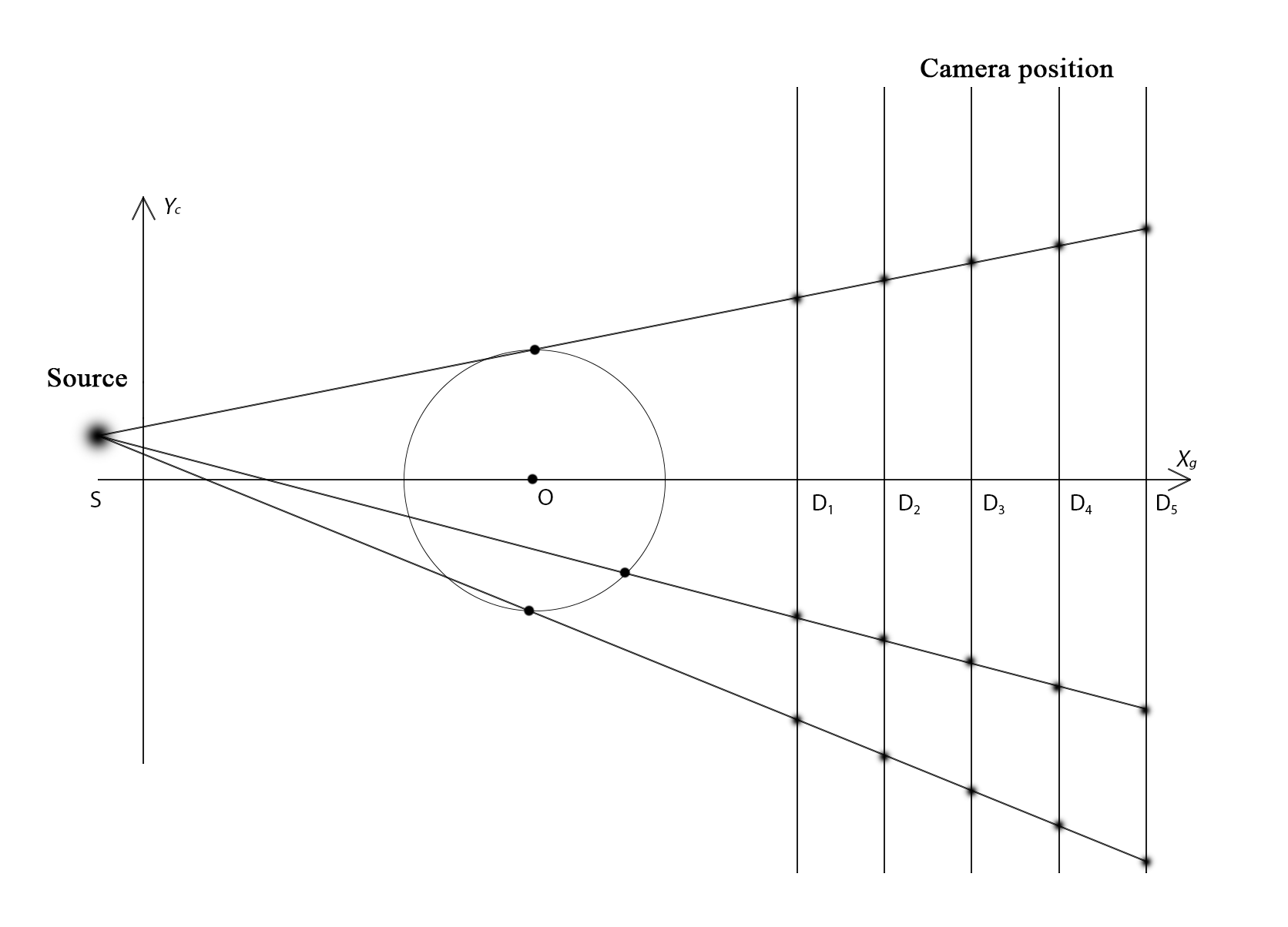}}
\caption{Determination of the source $Y_c$ position, and of the source to detector distance SDD}
\label{fig:calib1}
\end{figure}

To determine the source position along the $Y_c$ axis, the measurement scheme shown in Figure~\ref{fig:calib1} can be used.
One needs to perform several sets of measurements of the shadow coordinate obtained for different "cross" height positions by moving the camera towards and away from the source. Different "cross" height positions are achieved by simply rotating the phantom by various angles.

\begin{figure}[H] 
\center{\includegraphics[width=0.95\linewidth]{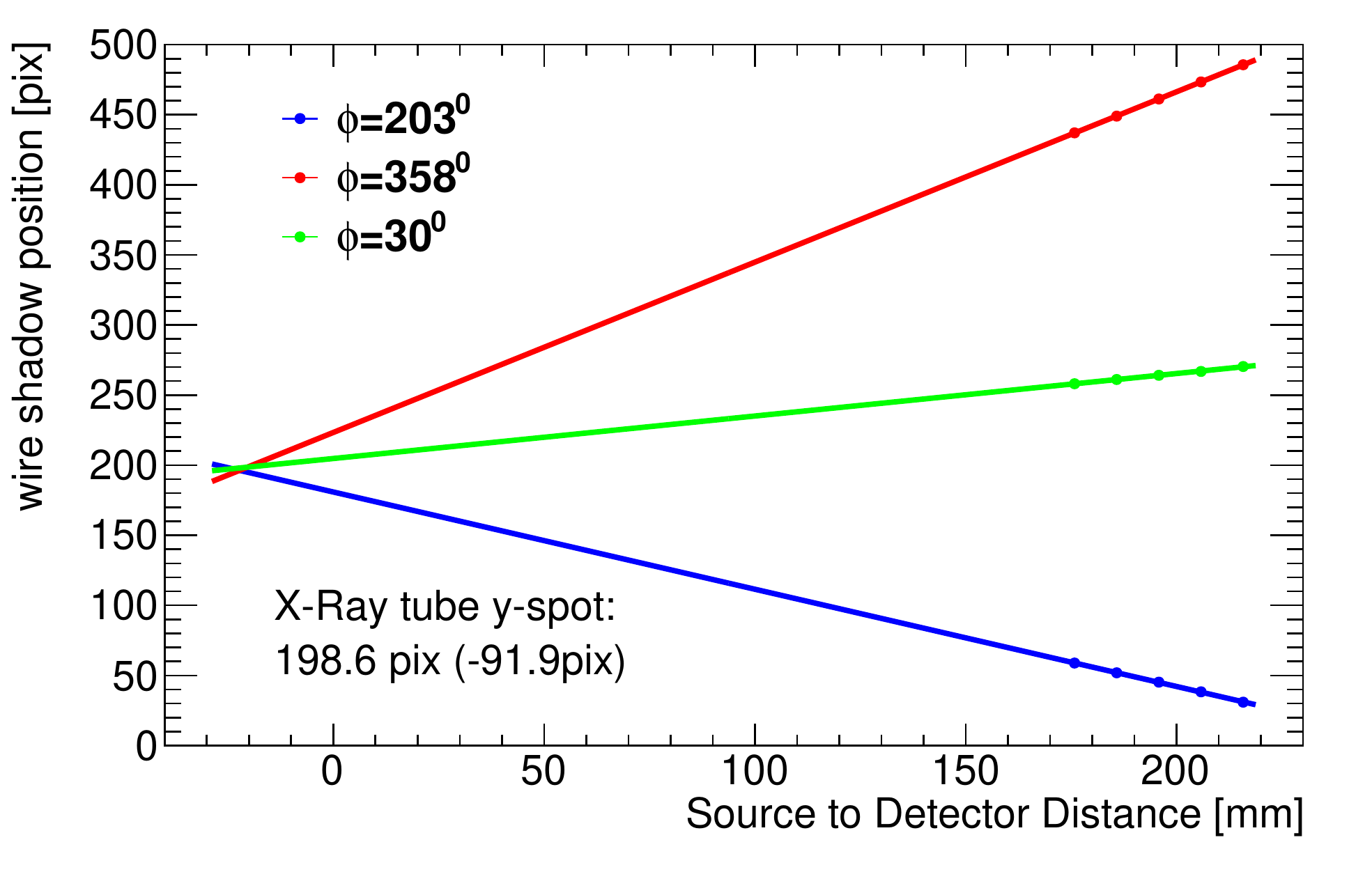}}
\caption{Results of calculation of the source $Y_c$ position and correction of the SDD distance between the source and the camera}
\label{fig:calib_res1}
\end{figure}

In our case, the $Y_c$ coordinate of the shadow cross have been measured for three positions of the gantry rotation angle in five camera positions along the $X_g$ axis connecting the camera to the source. 
In Figure~\ref{fig:calib_res1} the green line corresponds to the shadow position measurements at 30$^\circ$ rotation angle, blue line corresponds to 203$^\circ$ gantry rotation angle and the red line corresponds to 358$^\circ$ gantry rotation angle. All three lines have to cross at the same point, which is the position of the x-ray source along the camera $Y_c$ axis.
As it is seen from the graph, this method is suitable not only for determining of the source $Y_c$ position, but also to determine the distance SDD from the source focal spot to the camera, which in this case turned out to be approximately two centimeters greater than the nominal.

\section{The impact of the tomograph alignment to the image reconstruction quality}

\begin{figure}[H] 
\center{\includegraphics[width=0.75\linewidth]{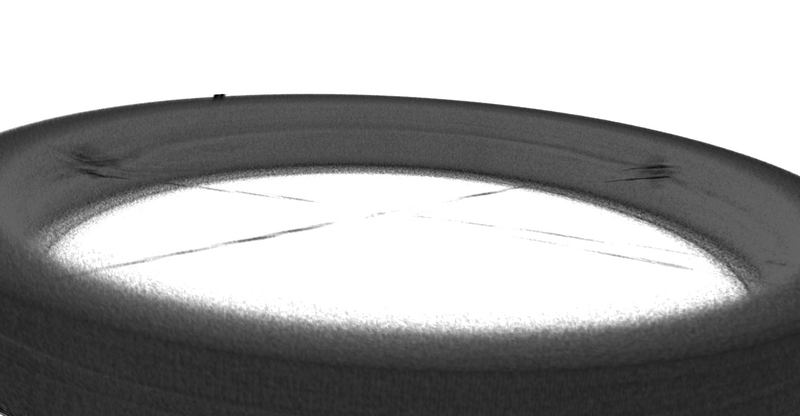}}
\caption{Phantom image reconstructed without corrections}
\label{fig:rec1}
\end{figure}

\begin{figure}[H] 
\center{\includegraphics[width=0.75\linewidth]{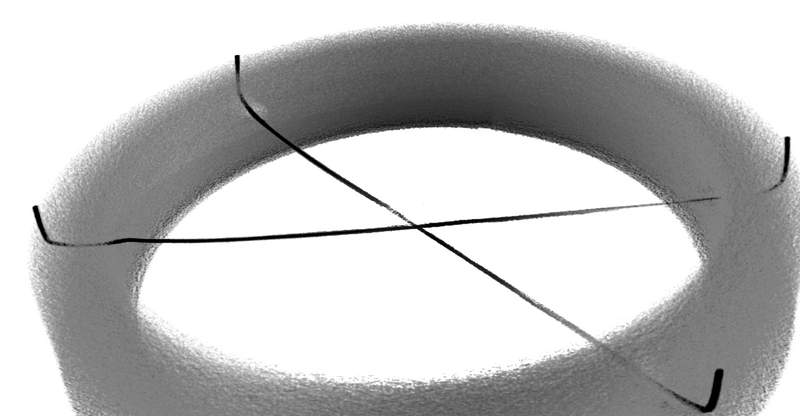}}
\caption{Phantom image reconstructed with the alignment corrections applied}
\label{fig:rec2}
\end{figure}

Geometrical corrections obtained in the procedure of the tomograph alignment can be used as input parameters for the image reconstruction. The plastic phantom with two additional wires strung in its cross section has been used as a scanned sample.
In the first case, the image reconstruction is performed using the nominal parameters of the geometry ({Figure~\ref{fig:rec1}}). In the second case, corrections obtained in the procedure of the tomograph alignment have been applied ({Figure~\ref{fig:rec2}}).

From these figures it is evident that without geometrical corrections the reconstruction is of lower quality, with several artifacts. In particular, the wires on the reconstructed image are look duplicated and dashed. The implementation of the alignment corrections results in a significant improvement in the quality of the reconstructed image.

\section{Spatial resolution}

The overall spatial resolution of the tomograph depends both on the quality of the original projections and also on the quality of the reconstruction. In the first case, it is influenced by the physical processes in the material of the detector, the detector pixel size, the size of the x-ray tube focal spot, and the zoom factor. Such factors as mechanical precision of the movement of the tomograph parts and the precise alignment of the system also affect the quality of the image reconstruction. In this section several methods are described that were used to obtain a quantitative estimation of the overall spatial resolution of the system.

\begin{figure}[H] 
\center{\includegraphics[width=0.95\linewidth]{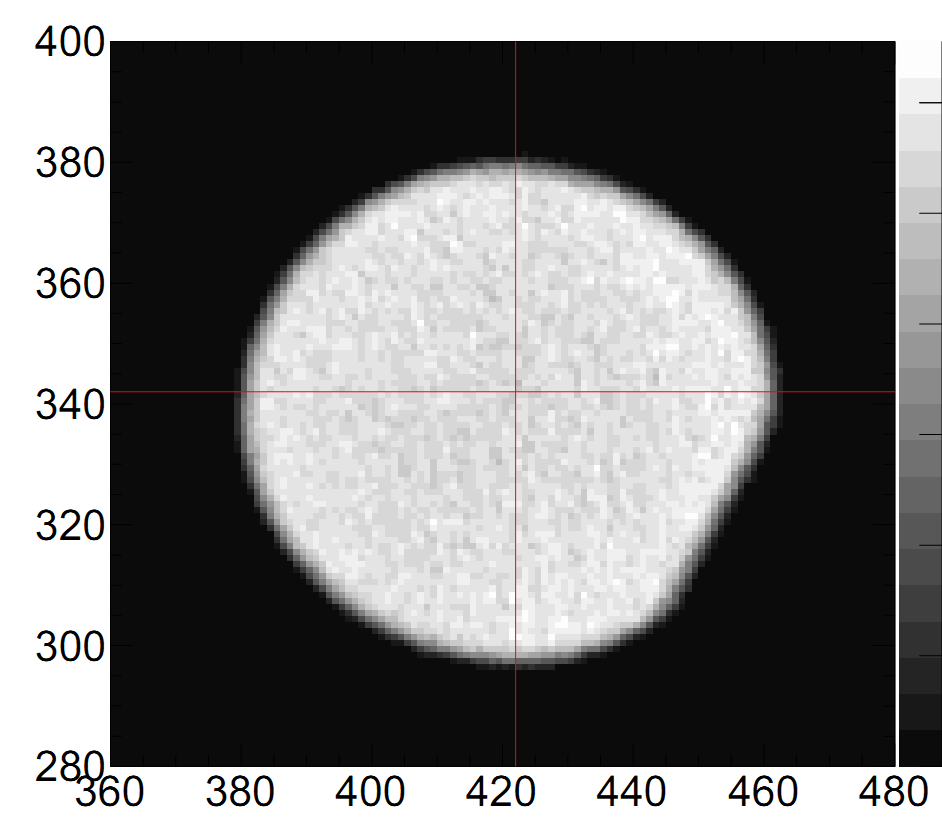}}
\caption{Sum over 20 reconstructed slices of aluminum wire (the wire is slightly damaged from the one side).}
\label{fig:al_wire_1}
\end{figure}

\begin{figure}[H] 
\center{\includegraphics[width=0.95\linewidth]{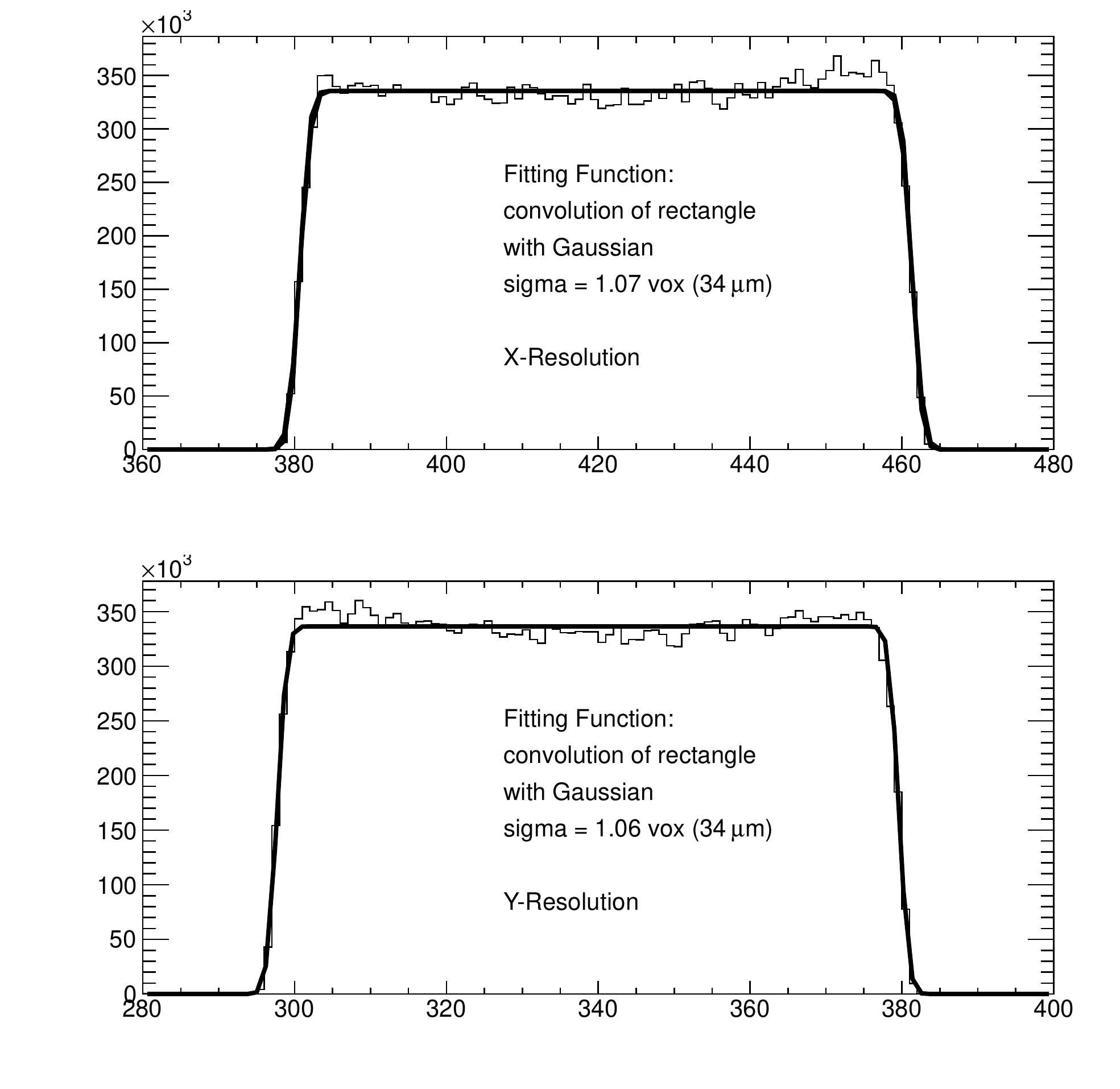}}
\caption{Approximation of the wire cross-section using convolution of a Gaussian with a rectangular function along the $X_s$ axis (top plot) and along the $Y_s$ axis (bottom plot).}
\label{fig:al_wire_2}
\end{figure}

The first method is based on indirect estimation of the spatial resolution using a Gaussian approximation to the point spread function (PSF). Aluminum wire of the diameter about 2.7~mm has been used as the scanned object.
To evaluate the spatial resolution of the tomograph, an image presenting the sum over 20 reconstructed slices of the wire is used (Figure~\ref{fig:al_wire_1}). A narrow area along the two axes presented by the red lines in the figure, which can be considered as rectangular, is selected for fitting. A function which is a convolution of a rectangle with a Gaussian is used here as an approximation function (Figure~\ref{fig:al_wire_2}). The rectangular function is a non-smeared initial image of the wire, while the Gaussian function is the system response to a point-like object. As a result of the fit, the value of $\sigma=34~\mu$m for the Gaussian point spread function (PSF) is determined along both axes. (The reader may note that Figure~\ref{fig:al_wire_2} has some beam hardening visible as a cupping artefact, but this does not affect our results.)

\begin{figure}[H] 
\center{\includegraphics[width=0.95\linewidth]{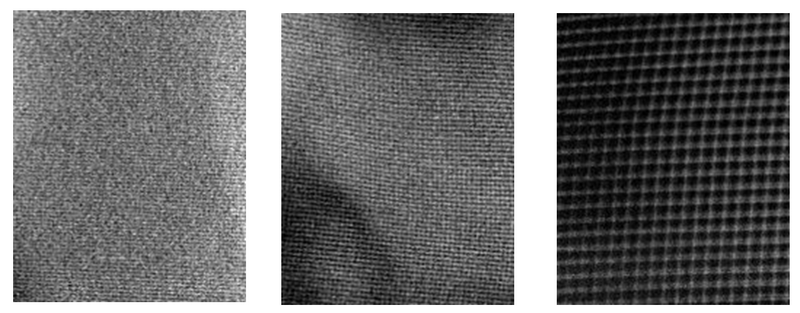}}
\caption{Reconstructed images of the three samples of stainless mesh: 25~$\mu$m diameter 67~$\mu$m repeat distance, 30~$\mu$m diameter 80~$\mu$m repeat distance and 50~$\mu$m diameter 265~$\mu$m repeat distance.} 
\label{fig:cloth1}
\end{figure}

To get a direct estimate of the spatial resolution, three different samples of stainless mesh have been scanned (Figure~\ref{fig:cloth1}). The first mesh has wires of 25~$\mu$m diameter and 67~$\mu$m repeat distance, the second mesh has wires of 30~$\mu$m diameter and 80~$\mu$m repeat distance and the third mesh has wires of 50~$\mu$m diameter and 265~$\mu$m repeat distance. Assuming that the response to a point object is Gaussian, the images of the two point-like objects become separated when the distance becomes larger than $2\sigma$ approximately. The smallest mesh is barely resolved, but for the second sample (30~$\mu$m diameter and 80~$\mu$m repeat distance) the mesh structure is fully visible. This is in a good agreement with the estimation above of the spatial resolution.

\begin{figure}[H] 
\center{\includegraphics[width=0.95\linewidth]{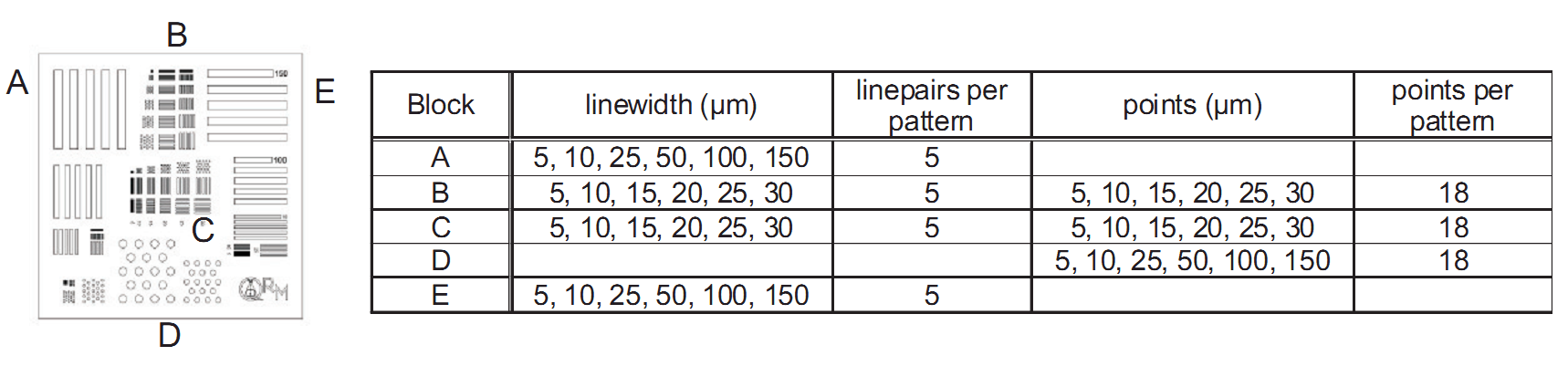}}
\caption{Description of the QRM-Barpattern-Phantom}
\label{fig:calibration_fantom_description}
\end{figure}

\begin{figure}[H] 
\center{\includegraphics[width=0.95\linewidth]{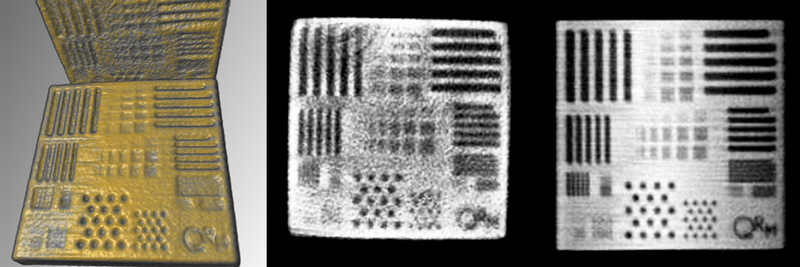}}
\caption{Reconstructed images of QRM-Barpattern-Phantom. Left plot: reconstructed 3D image. Middle plot: phantom cross-section in the X-Y plane. Right plot: phantom cross-section in the X-Z plane.}
\label{fig:calibration_fantom}
\end{figure}

A second independent direct check of a spatial resolution has been performed by scanning a standard calibration phantom, the QRM-Barpattern-Phantom (Figure~\ref{fig:calibration_fantom_description}). This calibration phantom consists of two perpendicular silicon 5x5~mm$^2$ chips with pattern bars and holes of 5 to 150 $\mu$m size and of 80 to 120~$\mu$m depth (Figure~\ref{fig:calibration_fantom_description}). Reconstructed images of the phantom are shown in Figure~\ref{fig:calibration_fantom}. The least size of visible bars on these pictures is 50 $\mu$m, this corresponds to 10 line-pairs per millimeter.

\section{Conclusions}

In computer tomography, a sufficiently accurate knowledge of the tomograph geometry is required to achieve good quality image reconstruction. In this paper, the precision of the movement of the MARS scanner parts has been studied and several geometrical parameters have been determined to improve the quality of the image reconstruction.
Several procedures for measuring the MARS scanner alignment have been developed and applied. The resulting corrections were used as input parameters for the image reconstruction algorithm.
The results of this work show that using these alignment corrections leads to a significant improvement in the quality of the image reconstruction. Using these corrections, several independent methods have been applied to make quantitative estimation of the final spatial resolution for the reconstructed image. The results of different methods are consistent and give an estimation for the tomograph spatial resolution of slightly greater than 30~microns.

\end{document}